\documentclass[pmlr]{jmlr}% new name PMLR (Proceedings of Machine Learning)

\RequirePackage{graphicx}
 \usepackage{booktabs}
\usepackage{longtable}% for long tables
\usepackage{adjustbox}
\usepackage{multirow}
\usepackage{array}
\usepackage{colortbl}
\usepackage{xcolor}

\makeatletter
\def\set@curr@file#1{\def\@curr@file{#1}} %temp workaround for 2019 latex release
\makeatother
\usepackage{siunitx} % newer version

\theorembodyfont{\upshape}
\theoremheaderfont{\scshape}
\theorempostheader{:}
\theoremsep{\newline}

\jmlrproceedings{PMLR}{Proceedings of Machine Learning Research}
\jmlrvolume{340}
\jmlryear{2026}
\jmlrworkshop{Machine Learning for Healthcare}

\title[Change Point--Aware Evaluation and Re-Calibration]{Change Point--Aware Evaluation and Re-Calibration of PPG-Based Blood Pressure Estimation}

\author{%
{\centering
  \Name{Yunwon Tae\nametag{$^{1,2,\ast}$}} \Email{ytae1@uci.edu}\\
  \Name{Minje Park\nametag{$^{1,\ast}$}} \Email{minje.park@vuno.co}\\
  \Name{Gyunho Rho\nametag{$^{1,3}$}} \Email{gyunho.rho@sbri.co.kr}\\
  \Name{Dongjoon Yoo\nametag{$^{4}$}} \Email{djinyoo@gmail.com}\\
  \Name{Sunghoon Joo\nametag{$^{1,\dagger}$}} \Email{sunghoon.joo@vuno.co}\\
  \addr $^{1}$VUNO Inc.\\
  $^{2}$University of California, Irvine\\
  $^{3}$Research Institute for Future Medicine, Samsung Medical Center\\
  $^{4}$
Department of Critical Care Medicine and Emergency Medicine, Inha University Hospital\\[2pt]
  \upshape $^{\ast}$Equal contribution.\quad $^{\dagger}$Corresponding author.% 
\endgraf}%
}

\begin{document}

\maketitle

\begin{abstract}
  Non-invasive continuous blood pressure (BP) monitoring using photoplethysmography (PPG) is a promising alternative to cuff-based measurements. However, existing PPG-based BP estimation studies predominantly rely on aggregated performance metrics (e.g., mean absolute error) computed over entire evaluation intervals, which can obscure model failures during rapid BP fluctuations and limit clinical relevance. In this work, we propose a fluctuation-aware evaluation framework for PPG-based BP estimation based on time-series change point detection. Instead of heuristic BP thresholding (e.g., $\Delta\mathrm{BP} > 10\mathrm{mmHg}$), we identify BP change points by capturing abrupt distributional shifts in BP trajectories and evaluate estimation performance specifically during these fluctuation periods. Our analysis shows that several state-of-the-art models exhibit substantial performance degradation around BP change points, and that periodic test-time calibration is insufficient to handle such dynamic BP variations. To address this limitation, we introduce a targeted re-calibration framework triggered by detected BP change points, improving robustness without modifying model architectures. To the best of our knowledge, this is the first systematic evaluation of PPG-based BP estimation from a BP change point perspective, highlighting the importance of fluctuation-aware evaluation and calibration for real-world continuous BP monitoring.
\end{abstract}
% {\let\thefootnote\relax\footnotetext{$*$ Equal contribution.}\par}
% {\let\thefootnote\relax\footnotetext{$\dag$ Corresponding author.}\par}

\section{Introduction}
\label{sec:intro}
Continuous and non-invasive blood pressure (BP) monitoring plays a critical role in early detection of hemodynamic instability, guiding treatment decisions, and improving patient outcomes across both inpatient and ambulatory settings~\citep{shen2025review, rastegar2020non, juri2018impact}. With the growing demand for cuffless monitoring, photoplethysmography (PPG) has emerged as one of the most accessible and scalable sensing modalities, enabling continuous BP estimation through wearable devices and bedside monitoring systems~\citep{mukkamala2025cuffless}. Recent advances in deep learning have further accelerated this progress, with numerous PPG-based BP estimation models demonstrating promising performance on public benchmarks and controlled experimental environments~\citep{saha2025pulse, mohammadi2025cuff, ma2024stp}.

From a methodological perspective, existing PPG-based BP estimation studies can largely be categorized into two research directions: \textit{calibration-free} approaches and \textit{calibration-based} approaches. Calibration-free methods aim to directly estimate BP from PPG signals without requiring subject-specific adjustment, focusing on learning generalized mappings between waveform morphology and BP values~\citep{moulaeifard2025generalizable, samimi2023ppg, slapnivcar2019blood}. In contrast, calibration-based methods leverage intermittent ground-truth BP measurements to correct model predictions, often improving subject-specific accuracy through periodic re-calibration~\citep{joung2023continuous, jo2025test}.

Despite their methodological differences, both calibration-free and calibration-based approaches share a fundamental limitation: \emph{they do not explicitly account for BP fluctuations}. BP is inherently dynamic and non-stationary, often exhibiting abrupt transitions driven by physiological stress, changes in posture, the effect of medication, or acute pathological events~\citep{mukkamala2025cuffless, yang2025cuffless, mukkamala2021evaluation}.

\begin{figure*}
  \begin{center}
      \includegraphics[width=0.9\textwidth]{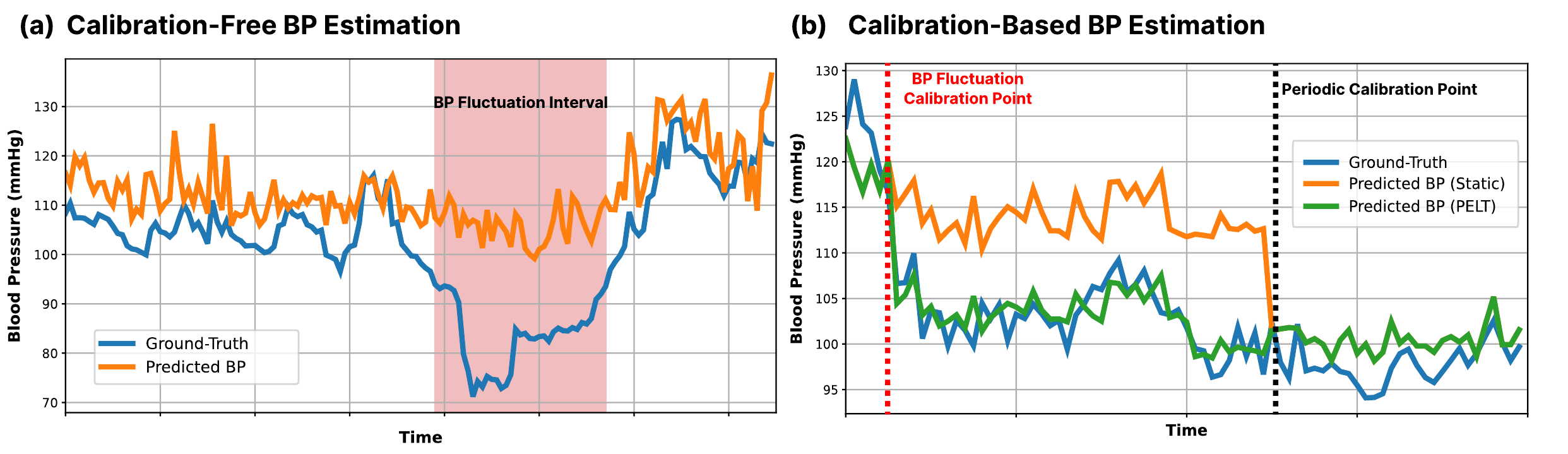}
      \vspace{-0.2in}
      \caption{Limitations of existing PPG-based blood pressure (BP) estimation approaches under BP fluctuations. (a) Calibration-free methods track the overall BP trend but fail to capture rapid fluctuations (red region), resulting in large errors during clinically critical periods. (b) Calibration-based methods with periodic re-calibration (orange) struggle when BP shifts occur between calibration points. Targeted re-calibration at BP fluctuation points (green) enables timely adaptation and maintains accurate tracking throughout.}
      \label{fig.intro-example}
      \vspace{-0.3in}
  \end{center}
  \end{figure*}

Figure~\ref{fig.intro-example} provides an intuitive illustration of these challenges. In Figure~\ref{fig.intro-example}(a), the calibration-free estimation appears to be accurate when assessed throughout the whole BP intervals; however, substantial prediction errors emerge during the BP fluctuation intervals. In Figure~\ref{fig.intro-example}(b), periodic calibration partially reduces long-term drift, yet fails to track sudden BP drops occurring between scheduled calibration points. Notably, when additional calibration is performed during fluctuation intervals, estimation trajectories more closely follow the ground-truth BP trend. These observations highlight that BP fluctuation awareness is essential for both evaluation and calibration design.

Motivated by this gap, our work systematically investigates the role of BP fluctuations in PPG-based BP estimation from both calibration-free and calibration-based perspectives. First, under the calibration-free setting, we explicitly evaluate model performance within fluctuation intervals and analyze whether simple strategies such as fluctuation-aware balanced sampling can improve robustness in these clinically critical regions. Second, under the calibration-based setting, we examine whether aligning calibration timing with BP fluctuation events---beyond fixed periodic schedules---can further enhance estimation accuracy.

A key challenge in enabling such analyses lies in defining BP fluctuations in a principled manner. Rather than relying on heuristic threshold rules (e.g., $\Delta\mathrm{BP} > 10\mathrm{mmHg}$)~\citep{wang2024delta,hong2024using}, we formulate BP instability through time-series change point detection. Specifically, we employ the Pruned Exact Linear Time (PELT) algorithm~\citep{killick2012optimal} to identify distributional transitions in continuous BP signals, providing a data-driven definition of fluctuation intervals. Using these detected change points, we conduct fluctuation-aware evaluation in calibration-free settings and design targeted re-calibration strategies in calibration-based settings. Our findings reveal that calibration effectiveness depends not only on frequency but critically on timing relative to physiological transitions.

However, directly applying change point detection in real-world deployment introduces practical constraints. Conventional algorithms such as PELT operate in an offline setting, requiring access to ground-truth BP signals and future observations, making real-time fluctuation detection infeasible. To bridge this gap, we further propose a framework that leverages PPG inputs to detect BP fluctuation events in an online manner. This enables real-time identification of re-calibration timing without relying on contemporaneous invasive BP measurements.

In summary, this work makes the following contributions: (1) We introduce a fluctuation-aware evaluation paradigm for PPG-based BP estimation, analyzing model behavior specifically within intervals exhibiting abrupt blood pressure changes. (2) We demonstrate that targeted re-calibration at BP change points significantly improves estimation accuracy over periodic re-calibration alone.
(3) We propose an online change point detection framework using only PPG signals, enabling targeted re-calibration without explicit BP monitoring.

To the best of our knowledge, this is the first study to investigate PPG-based BP estimation through the lens of BP change point dynamics, highlighting the necessity of fluctuation-aware evaluation and calibration for clinically reliable continuous BP monitoring.

\subsection*{Generalizable Insights about Machine Learning in the Context of Healthcare}
This paper offers the following generalizable insights for machine learning in healthcare:
\begin{itemize}
    \item Standard aggregate evaluation metrics (e.g., mean absolute error over entire recordings) can systematically mask model failures during clinically critical physiological transitions; fluctuation-aware evaluation protocols are essential for a reliable assessment of real-world performance.
    \item The timing of calibration matters at least as much as its frequency: targeted re-calibration at change points detected from PPG alone outperforms more frequent but temporally uninformed periodic re-calibration strategies.
    \item Online change point detection driven by self-supervised contrastive PPG representations provides a practical, architecture-agnostic trigger for adaptive calibration in continuous monitoring pipelines.
\end{itemize}

\section{Related Work}

\subsection{Calibration-Free PPG to BP Estimation}

Calibration-free PPG to BP estimation has traditionally relied on Pulse Wave Analysis (PWA) and Pulse Transit Time (PTT), where handcrafted morphological features or timing intervals are mapped to BP using conventional machine learning models~\citep{kachuee2015cuff}. More recently, deep learning architectures have been introduced to learn representations directly from raw PPG signals, demonstrating competitive estimation performance on benchmark datasets~\citep{moulaeifard2025generalizable, samimi2023ppg, slapnivcar2019blood}. Despite these advances, prior studies primarily report aggregate metrics computed over entire BP values. Such evaluation protocols may obscure model behavior during abrupt blood pressure fluctuations, where estimation reliability is clinically most critical. This highlights the need for fluctuation-aware analysis to better understand model robustness.

\subsection{Calibration-Based PPG to BP Estimation}

Calibration-based PPG to BP estimation approaches incorporate subject-specific references to personalize BP predictions~\citep{joung2023continuous, jo2025test}. These methods typically leverage periodic cuff-based measurements to adjust model outputs and compensate for inter-individual variability, improving estimation accuracy in real-world deployment settings. However, existing calibration paradigms commonly rely on fixed periodic schedules, implicitly assuming stable physiological conditions between calibration points~\citep{vybornova2021blood}. Given the inherently dynamic nature of blood pressure, such strategies may fail to provide timely correction during abrupt fluctuation intervals. Determining when calibration should be performed, therefore, remains underexplored, which we investigate in this work.

\subsection{BP Change Point Detection}

While explicit change point detection has rarely been explored in the context of BP estimation, several prior studies have addressed the concept of BP change from a different perspective. They primarily focus on tracking relative fluctuations rather than estimating absolute values~\citep{wang2024delta,hong2024using}, typically defining BP change as an event where the magnitude of variation over a specific time window exceeds a fixed threshold (e.g., $\Delta\mathrm{BP} > 10\mathrm{mmHg}$). However, their primary objective is to bypass the difficulty of regression, rather than to enable fluctuation-aware evaluation and determine optimal timing for re-calibration. Furthermore, relying on fixed absolute thresholds fails to account for inter-subject variability and neglects the temporal structure of BP sequences. In contrast, we approach identifying these transitions through time-series change point detection, enabling a statistically grounded criterion for re-calibration.

\subsection{Change Point Detection Algorithms in a Time-Series Domain}

Change Point Detection (CPD) is the process of partitioning a time-series sequence into distinct, statistically homogeneous segments by identifying abrupt shifts in the underlying distribution. CPD algorithms are categorized into offline~\citep{killick2012optimal,gharghabi2018domain} and online~\citep{adams2007bayesian,gharghabi2018domain} settings. Offline CPD retrospectively segments the entire sequence using full historical data. In contrast, online CPD detects changes in real-time using only historical data, making it suitable for streaming applications.

For re-calibration in continuous BP monitoring, online CPD is necessary to trigger re-calibration in real-time when BP changes occur. However, in the deployment setting, the reference BP cannot be obtained, making existing online BP change point detection methods inapplicable. These methods fundamentally rely on ground-truth BP measurements to identify distributional shifts. Therefore, we need a CPD method that works with PPGs alone, without requiring BP values as input.

\section{Problem Formulation}

\subsection{Calibration-Free PPG to BP Estimation}
The standard PPG to BP estimation is defined as a regression task that predicts systolic (SBP) and diastolic (DBP) blood pressure using a fixed-length PPG segment. Let $\mathrm{x}\in\mathbb{R}^{L}$ denote the normalized PPG input of length $L$, and $y\in\mathbb{R}^{+}$ denote the corresponding scalar blood pressure value. The estimation model is formulated as a function $\hat{y}=f(\mathrm{x};\theta)$, parameterized by $\theta$. The objective is to learn the optimal parameters $\theta^{*}$ that minimize the discrepancy between the prediction $\hat{y}$ and the ground-truth $y$, typically measured by the Mean Squared Error (MSE) loss.

\subsection{Blood Pressure Fluctuations and Change Points}\label{section.3.2}
Blood pressure is inherently non-stationary, exhibiting rapid fluctuations due to physiological stress or postural changes. Standard evaluation protocols, which aggregate errors over the entire recording, often mask significant performance degradation during these critical transitions. Models with low aggregated MAE can still produce clinically unacceptable errors during critical hemodynamic shifts.

To address this gap, we explicitly define \textbf{BP change points} as time steps where the underlying BP distribution undergoes a significant shift. Formally, given a BP sequence $y_{1:T}$, a change point $\tau$ is defined where the BP distribution $p(y_t)$ changes ($p(y_t \mid t < \tau) \neq p(y_t \mid t \geq \tau$). In this work, we derive ground-truth BP change points using the \textbf{PELT} algorithm~\citep{killick2012optimal} on the reference BP sequence.

Crucially, we define \textbf{Unstable BP Intervals} as the segments, derived from detected change points, exhibiting severe hypertensive or hypotensive levels (SBP $>$ 180 mmHg or mean arterial pressure $<$ 65 mmHg)~\citep{bress2024management,evans2021surviving}. To assess robustness under these clinically critical conditions, we introduce a dual-evaluation protocol measuring accuracy on both (1) all samples for baseline performance and (2) the samples within unstable intervals.

\subsection{Calibration-Based PPG to BP Estimation}
Beyond the temporal fluctuations described above, BP estimation faces the challenge of significant inter-subject variability.
Calibration-based estimation addresses this by incorporating subject-specific reference data. This task extends the input space to include a calibration set $\mathcal{D}_{cal}=\{(\mathrm{x}_{cal},y_{cal})\}$, acquired from the subject prior to the inference time $t$. The goal is to estimate the target blood pressure $y_{t}$ given the current input $\mathrm{x}_{t}$ and the calibration set $\mathcal{D}_{cal}$. We categorize existing approaches into two frameworks based on how they utilize $\mathcal{D}_{cal}$.

\paragraph{Input-Level Calibration} This framework utilizes the calibration data as an auxiliary input to the model, formulated as $\hat{y}_{t}=f(\mathrm{x}_{t},\mathrm{x}_{cal},y_{cal};\theta)$. For example, \textbf{PPG2BP-Net}~\citep{joung2023continuous} computes differential features between $\mathrm{x}_{t}$ and $\mathrm{x}_{cal}$, and combines them with calibration BP values $y_{cal}$ to regress the target value. In this setting, the parameters $\theta$ remain fixed during the inference phase.

\paragraph{Parameter-Level Calibration} This framework, exemplified by \textbf{Test-Time Calibration}~\citep{jo2025test}, formulates calibration as an online adaptation process. Instead of fixing the model parameters, the calibration set $\mathcal{D}_{cal}$ is used to dynamically update the parameters from an initial state $\theta$ to a personalized state $\theta'$. The inference is then performed as $\hat{y}_{t}=f(\mathrm{x}_{t};\theta')$, allowing the model to adapt its internal representations to the subject's physiological state.

\begin{figure*}[t]
    \centering
    \includegraphics[width=0.9\textwidth]{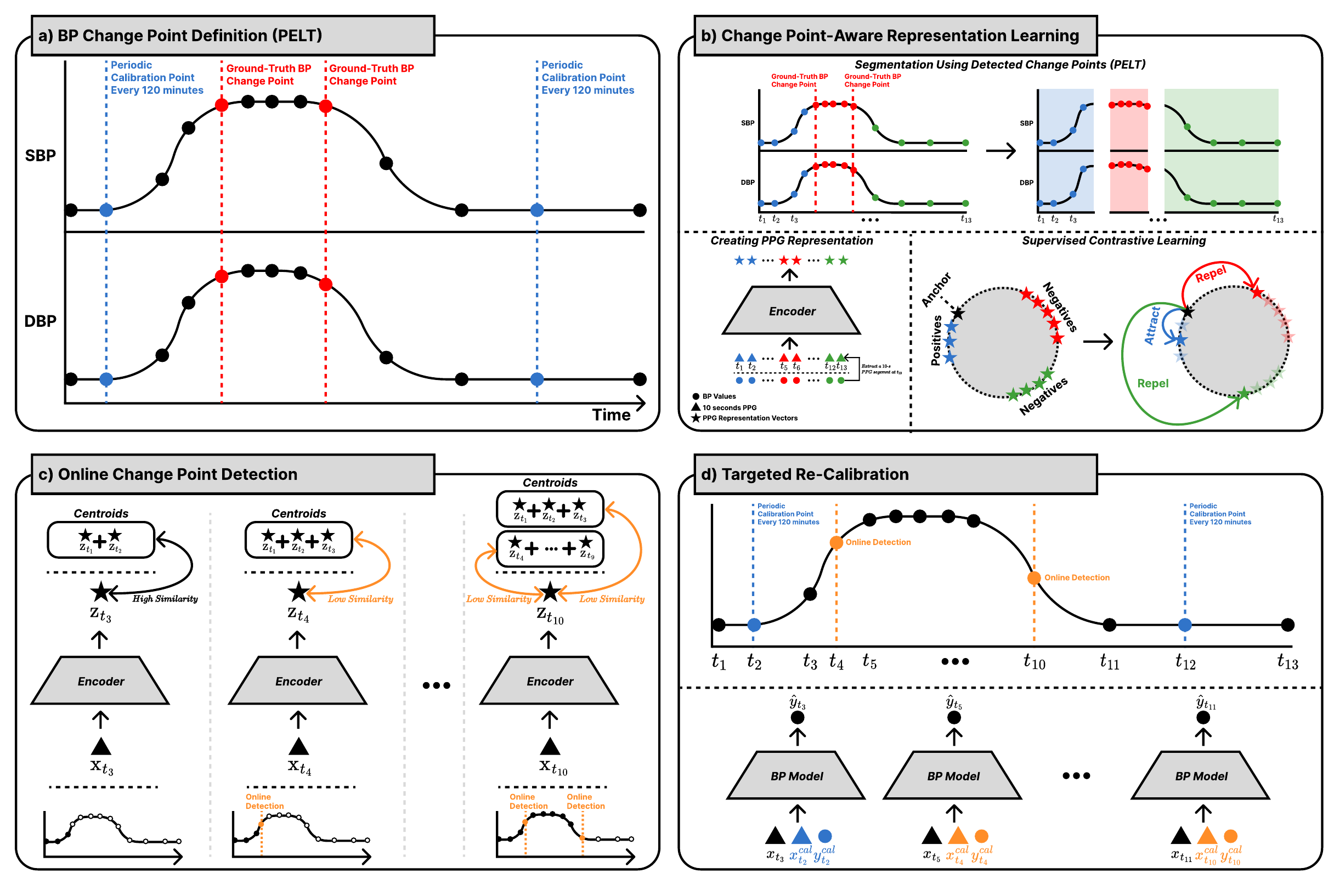}
    \vspace{-0.2in}
    \caption{Overview of the proposed targeted re-calibration framework. (a) Ground-truth blood pressure (BP) change points are obtained from the reference BP sequence. (b) Supervised contrastive learning trains a PPG encoder to attract features within the same segment (defined by change points) and repel those from different segments. (c) Online change point detection via sequential clustering: PPG features are sequentially compared against cluster centroids; low similarity to the current cluster triggers change point detection. (d) Detected change points supplement periodic calibration, enabling timely re-calibration upon physiological shifts.}
    \label{fig.algo_overview}
    \vspace{-0.2in}
\end{figure*}

\subsection{Targeted Re-Calibration}
Existing calibration methods rely on periodic re-calibration at fixed intervals. However, when BP fluctuates between calibration points, the model operates on outdated references, leading to degraded accuracy.

We propose \textbf{Targeted Re-Calibration}, which supplements periodic calibration with additional re-calibration triggered by detected BP change points mentioned in Section~\ref{section.3.2}. This approach ensures timely adaptation when physiological shifts occur while maintaining regular updates otherwise.

The challenge is that detecting BP change points typically requires continuous BP monitoring. In Section~\ref{section.4}, we address this by proposing an online change point detection framework that operates solely on PPG signals, enabling targeted re-calibration without explicit BP measurements.

\section{Targeted Re-Calibration with Online Blood Pressure Change Point Detection}\label{section.4}
To enable targeted re-calibration, we propose an online CPD framework based solely on PPG signals. The overall process is illustrated in \figureref{fig.algo_overview}. Our approach employs contrastive learning~\citep{chen2020simple} to learn discriminative PPG features, then applies sequential clustering to predict BP change points in real-time, eliminating the need for explicit BP tracking. The predicted change points serve as immediate triggers for re-calibration.

\begin{figure*}[t]
    \centering
    \includegraphics[width=0.9\textwidth]{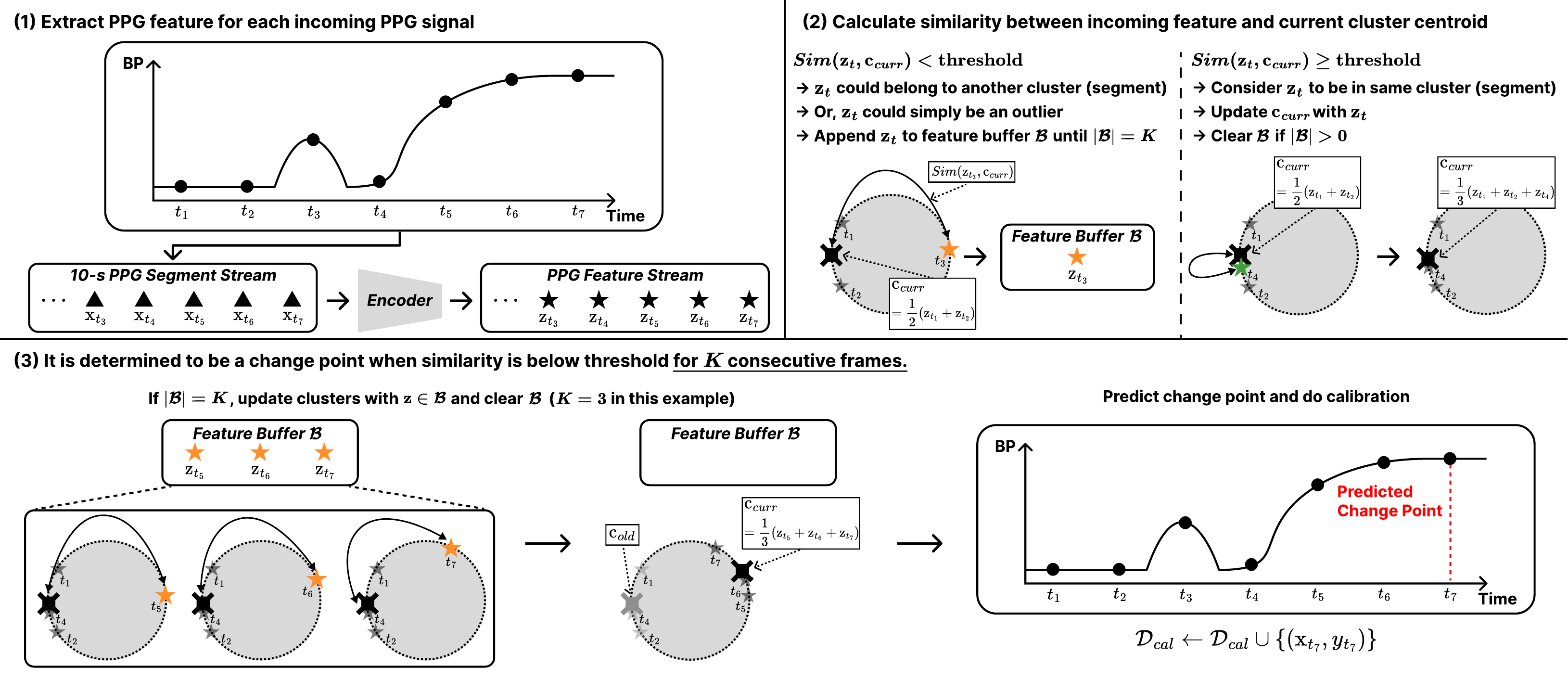}
    \vspace{-0.2in}
    \caption{An illustration of a sequential clustering algorithm for PPG-based blood pressure change point detection and re-calibration. (1) The PPG feature ($\mathrm{z}_{t}$) is extracted sequentially from the incoming PPG signal ($\mathrm{x}_t$). (2) Similarity between incoming features and the current cluster centroid ($\mathrm{c}_{curr}$) determines whether the physiological state changes. (3) The change point is detected when similarity remains below the threshold for $K$ consecutive frames, triggering cluster update and re-calibration.}
    \label{fig.algo_overview2}
    \vspace{-0.2in}
\end{figure*}

\subsection{Supervised Contrastive Learning for Segment-Aware PPG Representation}\label{section.4.2.1}
To detect BP change points effectively without direct BP measurements, we require a PPG representation space where distinct physiological states are separable. We achieve this by training a PPG feature encoder using \textbf{Supervised Contrastive Learning (SCL)}~\citep{khosla2020supervised}, guided by reference BP change points.

Let $\mathrm{x}_{1:T}=(\mathrm{x}_1, \mathrm{x}_2, \ldots,\mathrm{x}_T)$ denote the sequence of input PPG signals.
We define an encoder $h(\cdot;\phi)$ that maps each PPG signal $\mathrm{x}_t$ to a normalized feature embedding $\mathrm{z}_t=h(\mathrm{x}_t;\phi)\in\mathbb{R}^d$, where $d$ denotes the feature dimension.
Assume we are given a set of ground-truth BP change points $\mathcal{T}=\{\tau_1,\tau_2,...,\tau_M\}$. For each time step $t$ that falls in the segment $[\tau_m,\tau_{m+1})$, we define the set of positive indices $P(t)=\left\{ p \, | \, \tau_{m}\leq p <\tau_{m+1},\,p\neq t \right\}$, meaning all other time steps that belong to the same segment. The encoder is then optimized using the supervised contrastive loss:
\[
    \ell_{SCL}(t)=\frac{-1}{|P(t)|}\sum_{p\in P(t)}\log\frac{\exp(\mathrm{z}_{t}\cdot\mathrm{z}_{p}/\gamma)}{\sum_{a\in A(t)}\exp(\mathrm{z}_{t}\cdot\mathrm{z}_{a}/\gamma)},
\]
where $A(t)=\{a\,|\,1\leq a\leq T,\,a\neq t\}$ and $\gamma$ is a scalar temperature parameter. By minimizing this objective, the encoder learns to cluster PPG samples from the same segment while pushing apart those from different segments, ensuring that BP fluctuations result in distinct shifts in the feature space. Once trained, this encoder is deployed for online CPD.

\subsection{Online Change Point Detection via Sequential Clustering}\label{section.4.2.2}
To identify BP change points in real-time, we employ a \textbf{Sequential Clustering} algorithm on the feature stream $\mathrm{z}_{1:T}$. The algorithm maintains a dynamic set of cluster centroids $\mathcal{C}$, tracking an \textbf{active centroid} $\boldsymbol{\mathrm{c}_{curr}}$ that represents the current physiological state.

The core logic operates by sequentially checking the cosine similarity between the incoming feature $\mathrm{z}_{t}$ and the active centroid ($Sim(\mathrm{z}_{t},\mathrm{c}_{curr})$). If $\mathrm{z}_{t}$ remains sufficiently close to the active centroid with a threshold $\eta$ ($Sim(\mathrm{z}_{t},\mathrm{c}_{curr})\geq\eta$), we update $\mathrm{c}_{curr}$ via incremental averaging:
\begin{equation}
    \label{eq:update-c-curr}
    \mathrm{c}_{curr}\leftarrow\frac{N_{curr}\cdot \mathrm{c}_{curr}+\mathrm{z}_t}{N_{curr}+1},\,N_{curr}\leftarrow N_{curr}+1
\end{equation}
where $N_{curr}$ is the number of samples accumulated in the current state.

However, when a significant deviation occurs ($Sim(\mathrm{z}_t,\mathrm{c}_{curr})<\eta$), the algorithm searches for the centroid with the highest similarity to $\mathrm{z}_t$ among all previously stored centroids, denoted $\mathrm{c}^*$. If a match is found ($Sim(\mathrm{z}_t,\mathrm{c}^*)\geq\eta$), the algorithm transitions to that existing state, updating the active centroid accordingly:
\begin{equation}
    \label{eq:update-c-prev}
    \mathrm{c}_{curr}\leftarrow\frac{N^*\cdot\mathrm{c}^*+\mathrm{z}_t}{N^*+1},\,N_{curr}\leftarrow N^*+1
\end{equation}
where $N^*$ is the number of samples accumulated in the matched state $\mathrm{c}^*$. Otherwise, a new centroid is initialized with $\mathrm{z}_t$:
\begin{equation}
    \label{eq:new-centroid}
    \mathrm{c}_{curr}\leftarrow\mathrm{z}_t,\,N_{curr}\leftarrow 1,\,\,\mathcal{C}\leftarrow\mathcal{C}\cup\{\mathrm{z}_t\}.
\end{equation}
Any such transition flags a change point and triggers re-calibration. The overview is illustrated in \figureref{fig.algo_overview2}.

\subsection{Practical Refinements for Robust Change Point Detection}
The sequential clustering algorithm described in Section~\ref{section.4.2.2} assumes a fixed threshold $\eta$ and triggers a state transition immediately upon detecting a deviation. In practice, this naive approach is prone to false positives: even a single outlier can exceed the threshold, and the PPG features tend to be unstable right after a genuine state change. Raising the threshold mitigates this but risks missing subtle shifts. To address these issues, we incorporate two refinements.

\paragraph{Duration-Dependent Thresholding}
The fixed threshold $\eta$ in Section~\ref{section.4.2.2} is replaced with a duration-dependent variant $\eta=\eta_{base}\cdot(1-\exp({-\lambda n}))$, where $n$ denotes the duration of the current segment, measured as the number of frames elapsed since the last detected change point. Here, $\eta_{base}$ denotes the target threshold and $\lambda$ controls how quickly the threshold reaches it.
The threshold starts low right after a state transition to prevent false change points when the features are still unstable, and gradually tightens to $\eta_{base}$ as the state stabilizes.

\paragraph{Delayed Prediction}
To filter out short-term outliers, we enforce a persistence constraint. When a deviation occurs, $\mathrm{z}_{t}$ is held in a feature buffer $\mathcal{B}$ rather than triggering an immediate transition. If no deviation is detected, $\mathcal{B}$ is reset. A change point is confirmed only if the deviation persists for $K$ consecutive frames (i.e., $|\mathcal{B}|=K$). Upon confirmation, clusters are updated with the buffered features following \equationref{eq:update-c-prev,eq:new-centroid}.

We present a more detailed procedure in \algorithmref{alg:online-cpd}, which can be found in Appendix~\ref{apd:third}.

\begin{table}[t]
\centering
\small
\caption{PulseDB Dataset Statistics for MIMIC-III and VitalDB.}
\vspace{0.3cm}
\begin{tabular}{lcccc}
  \toprule
  \multirow{2}{*}{\textbf{}} &
  \multicolumn{2}{c}{\textbf{MIMIC-III}} &
  \multicolumn{2}{c}{\textbf{VitalDB}} \\
  \cmidrule(lr){2-3} \cmidrule(lr){4-5}
  & \textbf{Train} & \textbf{Test} & \textbf{Train} & \textbf{Test} \\
  \midrule
  N. Subjects/Cases & 1,213/2,095 & 135/236 & 1,293/1,347 & 144/146 \\
  N. 10s Segments       & 436,680     & 54,000  & 465,480     & 57,600  \\
 N. Unstable Segments& 53,518& 6,085& 54,507&7,132\\
  SBP (mean ± SD)          & 121.94 ± 22.58 & 122.47 ± 21.83 & 115.48 ± 18.93 & 115.43 ± 18.72 \\
  DBP (mean ± SD)          & 60.73 ± 13.14  & 60.91 ± 12.58 & 62.92 ± 12.08  & 63.02 ± 11.87 \\
  \bottomrule
\end{tabular}
\label{table.one}
\end{table}

\section{Cohort}

\subsection{Dataset}\label{sec:dataset}
We conduct our experiments using the publicly available PulseDB dataset~\citep{wang2023pulsedb}, a large-scale benchmark for cuff-less blood pressure (BP) estimation. PulseDB integrates waveform data from the MIMIC-III~\citep{johnson2016mimic} and VitalDB~\citep{lee2022vitaldb} datasets, comprising electrocardiogram (ECG), photoplethysmogram (PPG), and arterial blood pressure (ABP) signals.

The dataset consists of 10-second waveform segments sampled at 125\,Hz. To ensure high signal reliability, PulseDB applies strict quality control criteria: segments without complete cardiac cycles are excluded, PPG segments with negative skewness are removed, and samples with a correlation lower than 0.9 between PPG and ABP are discarded. As a result of this filtering, the retained segments exhibit irregular temporal spacing across subjects. Ground-truth systolic and diastolic blood pressure (SBP and DBP) values are derived directly from the ABP waveform.

We use split dataset provided from the PulseDB dataset, which is subject-independent training and test sets for both MIMIC-III and VitalDB sources. Table~\ref{table.one} summarizes the dataset statistics, including the number of subjects and cases, total number of 10-second segments, number of segments of unstable BP interval, and the distribution of SBP and DBP values across splits.

\subsection{Preprocessing}
All PPG signals are processed using a band-pass filter to suppress baseline drift and high-frequency noise. For experiments that leverage the Pulse-PPG pretraining model~\citep{saha2025pulse}, the filtered signals are resampled to 50~Hz to ensure compatibility with the pretraining configuration. In all other experimental settings, the original sampling rate of 125~Hz provided by the PulseDB dataset is retained. Before being fed into the models, each PPG segment is standardized using z-normalization, such that the input has zero mean and unit variance. This preprocessing pipeline is applied consistently across all experiments.

\subsection{Baselines}
We consider both calibration-free and calibration-based BP estimation baselines to comprehensively evaluate the proposed framework. We additionally include multiple BP change point definitions in the calibration-based setting.

\paragraph{Calibration-Free BP Estimation}
We evaluate four calibration-free baselines: \textit{Inception1D}~\citep{ismail2020inceptiontime,moulaeifard2025generalizable}, \textit{PaPaGei-S}~\citep{pillai2025papagei}, \textit{Pulse-PPG}~\citep{saha2025pulse}, and \textit{AnyPPG}~\citep{nie2025anyppg}. Except for \textit{Inception1D}, which is trained from scratch, all models are initialized from pretrained checkpoints. All models are fine-tuned on PulseDB using the same training protocol for fair comparison. We adopt pretrained models to leverage robust and generalizable PPG representations, which have been shown to be effective in handling signal variability and distribution shifts. Such properties are particularly important for evaluating model performance under BP fluctuation scenarios. Furthermore, representation learning--based models are well suited for our setting, as the proposed framework relies solely on PPG signals to detect online BP change points without explicit BP measurements. A discriminative and stable PPG representation is therefore critical for reliable online change point detection.

\paragraph{Calibration-Based BP Estimation} For calibration-based baselines, we consider two representative calibration methods. First, we evaluate an input-level calibration approach using \textbf{PPG2BP-Net}~\citep{joung2023continuous}, where the model parameters are fixed during inference and calibration data are provided as additional inputs. Second, we adopt \textbf{Test-Time Calibration (TTC)}~\citep{jo2025test}, a parameter-level calibration method that dynamically updates model parameters at inference time using calibration samples. \textbf{TTC} enables personalized adaptation by optimizing model parameters based on subject-specific calibration data and serves as a strong baseline for dynamic BP estimation.

\paragraph{Different BP Change Point Definitions}
Periodic calibration is fixed at a 120-minute duration across all experiments, reflecting a clinically reasonable setting for ICU and operating room data (MIMIC and VitalDB), with additional ablations using shorter intervals reported in Appendix~\ref{apd:fourth}. In calibration-based experiments, we evaluate multiple BP change point definitions to disentangle the effect of re-calibration timing from re-calibration frequency. In addition to change points detected by \textit{PELT}, we consider a \textit{$\Delta$BP} criterion, where a change point is triggered when the absolute BP difference exceeds 10~mmHg, following prior work~\citep{wang2024delta}. We also include a \textit{Random} baseline that matches the number of change points detected by \textit{PELT} while randomly sampling their temporal locations.
This design controls for re-calibration frequency and isolates the importance of where re-calibration is applied.

\subsection{Training Details}
For calibration-based baselines, including \textbf{PPG2BP-Net} and \textbf{TTC}, we follow the original model architectures and hyperparameter settings. For \textbf{TTC}, although the original formulation leverages both ECG and PPG signals, we modify the input to use PPG only for consistency with the cuffless BP estimation setting. Calibration-free baselines are implemented following their original publications. Minor deviations in hyperparameter settings are introduced only for the BP estimation fine-tuning stage and the online CPD module; details are provided in Appendix~\ref{apd:second}.

\section{Results on Real Data} 

\begin{table}[t]
\centering
\caption{Comparison of BP estimation performance on whole time-series segments and clinically unstable BP intervals. Mean absolute error in mmHg with 95\% confidence intervals are reported. Baseline (Median) represents the result of predicting the training set median for all samples.}
  \begin{tabular}{lcccc}
  \toprule
  \multirow{2}{*}{\textbf{Models}} &
  \multicolumn{2}{c}{\textbf{Whole}} &
  \multicolumn{2}{c}{\textbf{Unstable BP Intervals}} \\
  \cmidrule(lr){2-3} \cmidrule(lr){4-5}
  & \textbf{SBP  ↓} & \textbf{DBP  ↓} & \textbf{SBP  ↓} & \textbf{DBP  ↓} \\
  \midrule
  Baseline (Median) & 16.48 ± 0.04& 9.75 ± 0.02& 21.10 ± 0.07& 11.85 ± 0.04\\
  \midrule
  \multicolumn{5}{c}{\textbf{Vanilla}} \\
  \midrule
  Inception1D       & 13.79 ± 0.03& 8.36 ± 0.02& 17.04 ± 0.07& 8.86 ± 0.04\\
  PaPaGei-S         & 13.68 ± 0.03& 8.66 ± 0.02& 15.81 ± 0.07& 9.12 ± 0.04\\
  Pulse-PPG         & 13.72 ± 0.03& 8.37 ± 0.02& 17.84 ± 0.07& 9.38 ± 0.04\\
  AnyPPG            & 14.79 ± 0.04& 8.41 ± 0.02& 17.40 ± 0.07& 10.03 ± 0.04\\
  \midrule
  \multicolumn{5}{c}{\textbf{Interval-balanced}} \\
  \midrule
  Inception1D       & 13.67 ± 0.03& 8.17 ± 0.02& 16.40 ± 0.07& 10.08 ± 0.04\\
  PaPaGei-S         & 13.82 ± 0.03& 8.70 ± 0.02& 15.34 ± 0.07& 8.38 ± 0.04\\
  Pulse-PPG         & 14.00 ± 0.03& 8.41 ± 0.02& 16.41 ± 0.07& 9.48 ± 0.04\\
  AnyPPG            & 13.92 ± 0.03& 8.41 ± 0.02& 17.32 ± 0.07& 10.30 ± 0.04\\
  \bottomrule
  \end{tabular}
\label{table.two}
\end{table}

\subsection{Fluctuation-Aware Evaluation}
Table~\ref{table.two} presents a fluctuation-aware evaluation of blood pressure (BP) estimation models, comparing performance over the entire recording (Whole) and over unstable BP intervals.

Across all models and training strategies, BP estimation errors consistently increase during unstable intervals compared to whole-interval evaluation. Specifically, we observe an average degradation of approximately 2--4~mmHg MAE for systolic BP (SBP) and 1--2~mmHg for diastolic BP (DBP). This trend highlights that standard evaluation over the full time-series substantially underestimates model errors during clinically important BP fluctuation periods.

Notably, models trained with the interval-balanced strategy (i.e., balanced sampling strategy) demonstrate improved robustness during unstable BP intervals compared to their vanilla counterparts. While whole-interval performance remains comparable, interval-balanced training consistently reduces SBP estimation error in unstable regions (e.g., from 17.04 to 16.40~mmHg for Inception1D and from 17.84 to 16.41~mmHg for Pulse-PPG). This suggests that explicitly balancing training samples across stable and unstable BP segments encourages models to better capture abrupt hemodynamic changes.

Overall, these results emphasize that accurate BP estimation during unstable intervals remains significantly more challenging than during stable periods, even for state-of-the-art PPG-based models. More importantly, they demonstrate the necessity of fluctuation-aware evaluation and training strategies to avoid overly optimistic performance assessments and to better reflect real-world clinical deployment scenarios.

\subsection{Importance of Re-Calibration\label{sec6.2}}
Calibration is a widely adopted strategy in PPG-based blood pressure (BP) estimation to mitigate inter-subject variability. However, most prior work~\citep{vybornova2021blood, nachman2025assessing} relies on fixed periodic calibration schemes (e.g., calibrating every week or month), implicitly assuming that BP dynamics evolve smoothly over time. Such approaches overlook a critical question: \emph{when} calibration is actually needed.

Table~\ref{table.three} investigates the impact of different re-calibration strategies under a fixed base calibration interval of 120 minutes. While periodic calibration provides a reasonable baseline, performance improves substantially when additional calibration points are introduced in a targeted manner. In particular, change point--aware re-calibration using \textit{PELT} yields the largest error reduction across both \textbf{TTC} and \textbf{PPG2BP-Net}, achieving the lowest SBP and DBP MAE.

Importantly, the results reveal that improved performance is not simply a consequence of increased calibration frequency. Although the \textit{$\Delta$BP} strategy introduces significantly more calibration points (17.36 points per case), its performance remains inferior to the \textit{PELT}-based approach, which uses nearly half as many calibration points (9.82 points per case). Similarly, random re-calibration, despite having a comparable number of calibration points, fails to consistently match the performance of the change point--driven strategy.

These findings highlight that indiscriminate or overly frequent calibration is neither efficient nor optimal. Instead, selectively triggering re-calibration during physiologically unstable periods enables models to adapt more effectively to abrupt BP shifts while minimizing unnecessary calibration overhead. This underscores the importance of \emph{fluctuation-aware} re-calibration strategies for reliable continuous BP monitoring in real-world settings.

\begin{table}[t]
\centering
  \caption{Impact of targeted re-calibration using ground-truth (offline) blood pressure change points. Mean absolute error in mmHg with 95\% confidence intervals is reported.   
  The bold represents the significant difference ($p < 0.05$) against other baseline methods. Points / Case denotes the average number of re-calibration points per case (mean ± std).
  The plus sign (``$+$'') keeps the 120-minute schedule as the base and adds the change points of strategies such as \textit{$\Delta$BP}, \textit{Random}, and \textit{PELT}. \textit{Random} uses the per-case point count of \textit{PELT} with uniformly sampled locations.
  Additional results for the unstable subgroup are provided in Appendix~\ref{apd:f2}.}
  \resizebox{\columnwidth}{!}{%
  \begin{tabular}{lccccc}
  \toprule
  \multirow{2}{*}{\textbf{Calibration Points}} &
  \multicolumn{2}{c}{\textbf{TTC}} &
  \multicolumn{2}{c}{\textbf{PPG2BP-Net}} &
  \multirow{2}{*}{\textbf{Points / Case}} \\
  \cmidrule(lr){2-3} \cmidrule(lr){4-5}
   &
  \textbf{SBP  ↓} & \textbf{DBP  ↓} &
  \textbf{SBP  ↓} & \textbf{DBP  ↓} & 
   \\
  \midrule
  Every 120 minutes          & 11.28 ± 0.06 & 6.22 ± 0.03 & 9.33 ± 0.05 & 5.21 ± 0.03 & 3.30 ± 1.52 \\ \midrule
  \hphantom{/}+ \textit{$\Delta$BP}       & 9.89 ± 0.05          & 5.34 ± 0.03          & 8.05 ± 0.05          & 4.53 ± 0.03         & 17.36 ± 22.57 \\
  \hphantom{/}+ \textit{Random}      & 9.27 ± 0.05          & 5.08 ± 0.03          & 7.44 ± 0.04          & 4.01 ± 0.03          & 9.91 ± 4.02 \\
  \hphantom{/}+ \textit{PELT}       & \textbf{8.97 ± 0.05} & \textbf{4.94 ± 0.03} & \textbf{5.84 ± 0.04} & \textbf{3.85 ± 0.02} & 9.82 ± 4.02 \\
  \bottomrule
  \end{tabular}}
\label{table.three}
\end{table}

\subsection{Online Change Point Detection}

Following the formulation described in Sections~\ref{section.4.2.1} and~\ref{section.4.2.2}, all models are trained to identify BP change points in an online manner using 10-second PPG segments as input.

As shown in Table~\ref{table.four}, we evaluate CPD performance using Area Under the Receiver Operating Characteristic Curve (AUROC), Area Under the Precision--Recall Curve (AUPRC), weighted accuracy~\citep{mosley2013balanced}, and Intersection of Union (IoU)~\citep{van2020evaluation}, reflecting the substantial class imbalance inherent to online BP change point detection, where fluctuation events are sparse. Among the evaluated models, Inception1D achieves the highest weighted accuracy (0.6457) and IoU (0.6149), while Pulse-PPG slightly outperforms others in terms of AUROC (0.7308). Overall, the evaluated models exhibit comparable detection performance across metrics.

Importantly, the objective of this experiment is not to establish a new state-of-the-art CPD method, but to assess the practical feasibility of integrating online BP change point detection as a trigger for targeted re-calibration. The results demonstrate that existing CPD models, when operated in an online setting, can reliably capture BP fluctuation intervals with sufficient accuracy to support downstream re-calibration strategies. This suggests that online CPD can serve as a viable and modular component within BP monitoring pipelines, and that future advances in CPD methodology may further enhance the effectiveness of fluctuation-aware re-calibration.

\begin{table}[t]
\centering
\caption{Online Change Point Detection Performance.}
    \begin{tabular}{lcccc}
    \toprule
    \textbf{Models} & \textbf{AUROC} & \textbf{AUPRC} & \textbf{Weighted Accuracy} & \textbf{IoU} \\
    \midrule
    Inception1D & 0.7275& \textbf{0.3079}& \textbf{0.6457}& \textbf{0.6149}\\
    PaPaGei-S & 0.7223& 0.3070& 0.6412& 0.6062\\
    Pulse-PPG & \textbf{0.7308}& 0.3066& 0.6112& 0.6099\\
    AnyPPG & 0.7286& 0.3075& 0.6176& 0.6106\\
    \bottomrule
    \end{tabular}
\label{table.four}
\end{table}

\subsection{Effectiveness of Online CPD-Based BP Re-Calibration}

Offline change point-based re-calibration has already demonstrated the importance of calibration timing over fixed or random strategies in Table~\ref{table.three}. Here, we focus on whether this benefit can be retained in an online setting, where BP change points must be detected sequentially without access to future information.

As shown in Table~\ref{table.five}, \textit{Online CPD}-based re-calibration consistently outperforms periodic and \textit{Random} strategies across both \textbf{TTC} and \textbf{PPG2BP-Net}. Despite relying solely on online change point detection, the proposed method achieves comparable or better accuracy while using a similar number of calibration points per case.

These results suggest that BP change points can be effectively identified and exploited in real time, enabling practical online re-calibration for continuous BP monitoring.

\begin{table}[t]
\centering
  \caption{Performance of targeted re-calibration using the proposed online change point detection. Mean absolute error in mmHg with 95\% confidence intervals is reported. The bold represents the significant difference ($p < 0.05$) against other baseline methods. Points / Case denotes the average number of re-calibration points per case (mean ± std).
  The plus sign (``$+$'') keeps the 120-minute schedule as the base and adds the change points of strategies such as \textit{Random} and \textit{Online CPD}. \textit{Random} uses the per-case point count of \textit{Online CPD} with uniformly sampled locations. Both \textit{Classification} and \textit{Online CPD} use \textit{Inception1D} as the CPD backbone.
  Additional results for the unstable subgroup are provided in Appendix~\ref{app:online-unstable}.}
  \resizebox{\columnwidth}{!}{%
  \begin{tabular}{lccccc}
  \toprule
  \multirow{2}{*}{\textbf{Calibration Points}} &
  \multicolumn{2}{c}{\textbf{TTC}} &
  \multicolumn{2}{c}{\textbf{PPG2BP-Net}} &
  \multirow{2}{*}{\textbf{Points / Case}} \\
  \cmidrule(lr){2-3} \cmidrule(lr){4-5}
   &
  \textbf{SBP  ↓} & \textbf{DBP  ↓} &
  \textbf{SBP  ↓} & \textbf{DBP  ↓} & 
   \\
  \midrule
  Every 120 minutes          & 11.28 ± 0.06 & 6.22 ± 0.03 & 9.33 ± 0.05 & 5.21 ± 0.03 & 3.30 ± 1.52 \\
  \midrule
  \hphantom{/}+ \textit{Random}& 9.52 ± 0.05& 5.20 ± 0.03& 7.60 ± 0.04& 4.07 ± 0.03
& 8.77 ± 3.34
\\
  \hphantom{/}+ \textit{Online CPD}  & \textbf{9.24 ± 0.05}& \textbf{5.12 ± 0.03}& 7.27 ± 0.04& 3.87 ± 0.02& 8.77 ± 3.33\\
  \bottomrule
  \end{tabular}}
\label{table.five}
\end{table}

\section{Discussion} 

\paragraph{Summary of findings}

In this work, we demonstrate that standard aggregate evaluation masks significant model failures during hemodynamic fluctuations. By introducing a fluctuation-aware protocol, we reveal consistent performance degradation in calibration-free models during these critical intervals---an average increase of 2--4 mmHg MAE for SBP and 1--2 mmHg for DBP---that would otherwise go unnoticed under conventional whole-recording metrics. This finding suggests that fluctuation-stratified reporting should serve as a standard complement to whole-interval evaluation, particularly when assessing model suitability for patients with highly dynamic BP profiles. For calibration-based approaches, we establish that the timing of re-calibration is at least as important as its frequency: change point-triggered re-calibration using \textit{PELT} outperforms both periodic-only and \textit{$\Delta$BP} strategies while using fewer calibration points, indicating that re-calibration design should prioritize alignment with physiological transitions rather than simply increasing update rates. Finally, by training a contrastive PPG encoder to detect these transitions online without access to ground-truth BP, we show that the benefit of targeted re-calibration can be retained in practical deployment, enabling timely re-calibration in ambulatory or step-down settings where invasive BP references are unavailable.

\paragraph{Why change point timing helps.}
The re-calibration triggers we compare differ in what each treats as a blood pressure fluctuation. The \textit{$\Delta$BP} criterion is a local, univariate rule that fires whenever the systolic change within a short window exceeds a fixed threshold, using SBP alone with no model of the underlying distribution, so it reacts to transient excursions and noise and yields redundant change points. \textit{PELT} instead detects distributional shifts in the joint SBP and DBP sequence, placing change points only where the signal statistics genuinely change, which is why its sparser triggers align more closely with true hemodynamic transitions.

\paragraph{Linking detection quality to re-calibration gain.}
The re-calibration gains in Table~\ref{table.five} depend on how well the online module detects true change points, not merely on triggering additional updates. To make this relationship explicit, we compare our formulation against a naive binary classifier trained on the same PELT-derived labels (Table~\ref{tab:cpd_w_cls}). Detection quality and downstream accuracy move together: raising AUROC from 0.5332 to 0.7275 lowers SBP MAE from 8.31 to 7.27 mmHg under the same \textbf{PPG2BP-Net} estimator, and the same ordering holds for the PaPaGei-S backbone. The consistency of this pattern across a from-scratch and a pretrained encoder indicates that the gain originates from the detection formulation rather than the choice of backbone.

\paragraph{What transitions the detector captures.}
The detected change points concentrate on a specific class of physiological event. Cross-referencing the unstable intervals with VitalDB vasopressor records (Table~\ref{tab:vasopressor}), the large majority of unstable segments (85.4\%) occur under vasopressor administration, indicating that PELT identifies pharmacologically driven hemodynamic shifts rather than arbitrary statistical fluctuations. These are also the events during which a stale calibration reference is most likely to mislead the estimator, which supports change point triggering as a clinically meaningful criterion.

\paragraph{Limitations}
The current framework relies on PulseDB data drawn from ICU and operating room settings; generalizability to ambulatory or wearable monitoring contexts remains to be validated. The online CPD module requires an initial training phase using ground-truth BP change points, which may limit applicability in fully unsupervised deployment scenarios. The PELT penalty hyperparameter was fixed across all subjects; subject-adaptive tuning may further improve change point localization. Finally, the targeted re-calibration strategy still assumes the availability of an occasional cuff measurement at detected change points, and further work is needed to explore fully cuffless adaptation mechanisms.

\bibliography{main}

\newpage
\appendix

\section{Further Implementation Details}\label{apd:first}
\subsection{Blood Pressure Change Point Detection with PELT}

We use the Pruned Exact Linear Time (PELT) algorithm to detect ground-truth BP change points from reference SBP and DBP sequences. In practice, we use the PELT implementation provided by the \texttt{ruptures}\footnote{\url{https://centre-borelli.github.io/ruptures-docs}} library. Table~\ref{tab:supple-pelt-config} summarizes the hyperparameter settings.

\begin{table}[h]
\centering
    \caption{Hyperparameters used for ground-truth BP change point detection using the PELT implementation in \texttt{ruptures}.}
    \begin{tabular}{lc}
        \toprule
        \textbf{Parameter} & \textbf{Value} \\
        \midrule
        \texttt{model} & \texttt{rbf} \\
        \texttt{min\_size} & 1 \\
        \texttt{jump} & 1 \\
        \texttt{pen} & 5 \\
        \bottomrule
    \end{tabular}
\label{tab:supple-pelt-config}
\end{table}

\subsection{Calibration-Free PPG to BP Estimation and Online Change Point Detection}\label{apd:second}
Table~\ref{tab:supple-encoder-config} summarizes the hyperparameters for PPG preprocessing, encoder training, and sequential clustering.
The encoder training configuration is shared for both calibration-free PPG to BP estimation and supervised contrastive learning for CPD.
For encoder backbones, we use Inception1D\footnote{\url{https://github.com/AI4HealthUOL/ppg-ood-generalization}}, PaPaGei-S\footnote{\url{https://github.com/Nokia-Bell-Labs/papagei-foundation-model}}, Pulse-PPG\footnote{\url{https://github.com/maxxu05/pulseppg}}, and AnyPPG\footnote{\url{https://github.com/Ngk03/AnyPPG}}, and adopt their original implementations for architecture-specific details.

\begin{table}[h]
\floatconts
  {tab:supple-encoder-config}
  {\caption{Hyperparameters for calibration-free BP estimation and online change point detection}}
  {
    \begin{tabular}{lc}
    \toprule
    \textbf{Parameter} & \textbf{Value} \\
    \midrule
    \multicolumn{2}{c}{\textbf{PPG Preprocessing}} \\
    \midrule
    Filter type& Third-order Butterworth \\
    Filter bandwidth& 0.5--8.0 Hz \\
    Normalization& Z-normalization \\
    \midrule
    \multicolumn{2}{c}{\textbf{PPG Encoder Training}} \\
    \midrule
    Epoch& 50 \\
    Batch size& 512 \\
    Optimizer& AdamW \\
    Learning rate& 0.001 \\
    Weight decay & 0.001 \\
    \midrule
    \multicolumn{2}{c}{\textbf{Sequential Clustering}} \\
    \midrule
    $\eta_{base}$ & 0.99 \\
    $\lambda$ & 0.05 \\
    $K$ & 5 \\
    \bottomrule
    \end{tabular}
  }

\end{table}

\subsection{Pseudo-Code for Online Change Point Detection via Sequential Clustering}\label{apd:third}
Algorithm~\ref{alg:online-cpd} summarizes the online change point detection procedure used in Section~\ref{section.4.2.2}, where change points are triggered by persistent deviations from the current cluster centroid under a duration-dependent similarity threshold. Further details can be found at the official implementation: \url{https://github.com/bakqui/ppg-bpcp}.

\begin{algorithm2e}[htbp]
\caption{Sequential clustering for online change point detection}
\label{alg:online-cpd}
\DontPrintSemicolon
\SetKwInOut{Initialize}{Initialize}

\KwIn{}
$\quad$ Feature stream $\mathrm{z}_{1:T}$\;
$\quad$ Threshold params $\eta_{base}, \lambda$\;
$\quad$ Persistence $K$\;
\KwOut{List of change points $\mathcal{T}$}

\Initialize{}

$\quad$$\mathcal{T} \leftarrow [1]$, $\mathcal{C} \leftarrow [\mathrm{z}_1]$\;
$\quad$$\mathrm{c}_{curr} \leftarrow \mathrm{z}_{1}$, $n \leftarrow 1$\;
$\quad$$k \leftarrow 0$, $\mathcal{B} \leftarrow []$\;
\BlankLine
\For{$t = 2$ \textbf{to} $T$}{
    $\eta_t \leftarrow \eta_{base} \cdot (1 - \exp(-\lambda n))$\;
    \BlankLine
    \eIf{$Sim(\mathrm{z}_t,\mathrm{c}_{curr}) \geq \eta_t$}
    {
        $k \leftarrow 0, \mathcal{B} \leftarrow []$\;
        $n \leftarrow n + 1$\;
        Update $\mathrm{c}_{curr}$ with $\mathrm{z}_{t}$ (\equationref{eq:update-c-curr})\;
    }
    {
        $k \leftarrow k + 1$\;
        Append $(\mathrm{z}_t,\eta_t)$ to $\mathcal{B}$\;
        \BlankLine
        \If{$k = K$}{
            \tcc{Change point detected. Update the current state.}
            Append $t$ to $\mathcal{T}$\;
            \BlankLine
            \tcc{Update centroids with stored features}
            \For{$(\mathrm{z},\eta) \in \mathcal{B}$}{
                $\mathrm{c}^* \leftarrow \operatorname{argmax}_{\mathrm{c}\in\mathcal{C}}\mathrm{Sim}(\mathrm{z},\mathrm{c})$\;
                \BlankLine
                \eIf{$Sim(\mathrm{z},\mathrm{c}^*) \geq \eta$}
                {
                    Update $\mathrm{c}_{curr}$ with $\mathrm{c}^*$ and $\mathrm{z}$ (\equationref{eq:update-c-prev})\;
                }
                {
                    Initialize new $\mathrm{c}_{curr}$ with $\mathrm{z}$ (\equationref{eq:new-centroid})\;
                }
            }
            \BlankLine
            $k \leftarrow 0, \mathcal{B} \leftarrow []$\;
            $n \leftarrow 0$\;
        }
    }
}
\Return $\mathcal{T}$\;
\end{algorithm2e}

\begin{figure*}[t]
    \centering
    \includegraphics[width=1\textwidth]{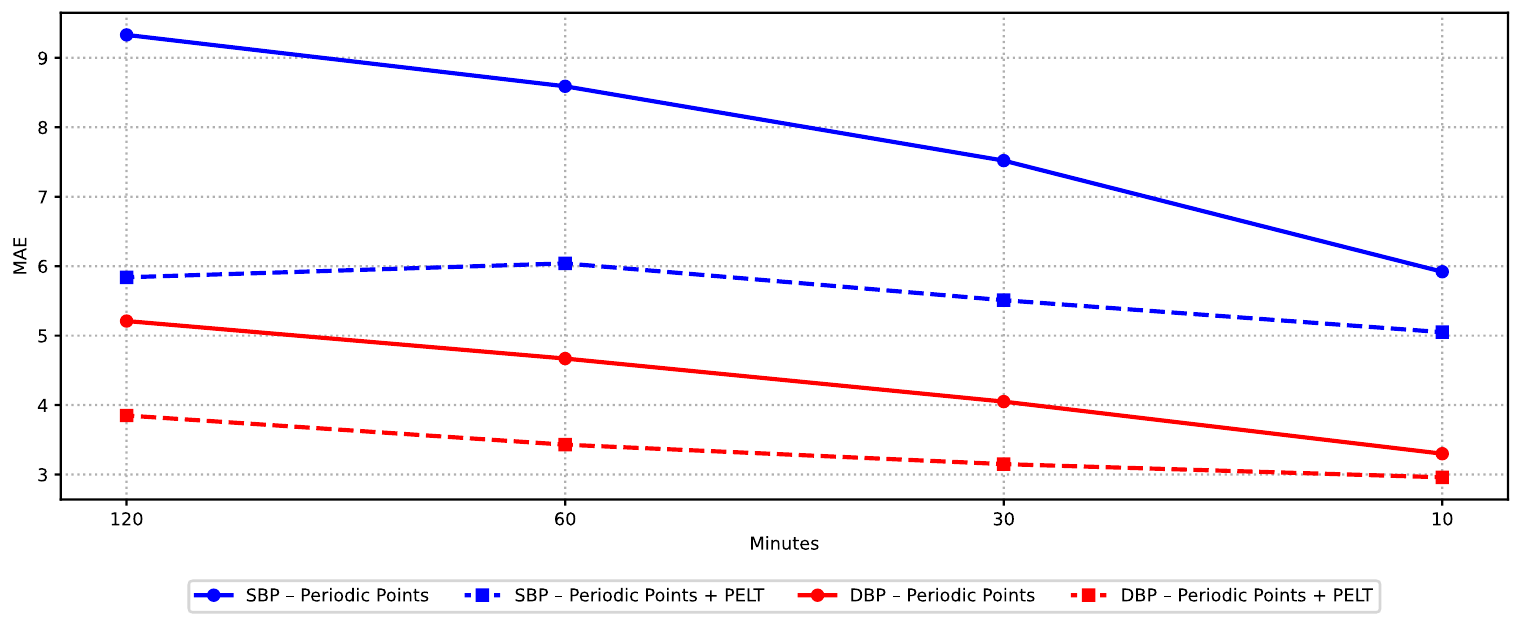}
    \vspace{-0.2in}
    \caption{Effect of periodic calibration interval on BP estimation error, with and without PELT-based change point triggering.}
    \label{fig.ablation}
\end{figure*}

\section{Effect of Periodic Calibration Frequency}\label{apd:fourth}
To analyze the effect of calibration frequency, we conduct an additional ablation study where the BP change point detector (PELT) is fixed, while the periodic calibration interval is varied among 120, 60, 30, and 10 minutes. Figure~\ref{fig.ablation} reports the resulting MAE for both systolic BP (SBP) and diastolic BP (DBP).

As the calibration interval becomes shorter, periodic calibration alone consistently improves estimation accuracy, indicating that more frequent re-calibration helps mitigate performance degradation caused by physiological drift. However, incorporating PELT-based change point triggering further improves performance across all calibration intervals. Notably, the relative gain from PELT remains evident even under short periodic intervals (e.g., 30 and 10 minutes), suggesting that re-calibration timing aligned with BP fluctuation events provides complementary benefits beyond increasing calibration frequency.

These results indicate that calibration effectiveness depends not only on how often re-calibration is performed, but also on when it is applied, reinforcing the importance of fluctuation-aware re-calibration strategies.

\section{Qualitative Results}\label{apd:fifth}
\begin{figure*}[t]
    \centering
    \includegraphics[width=1\textwidth]{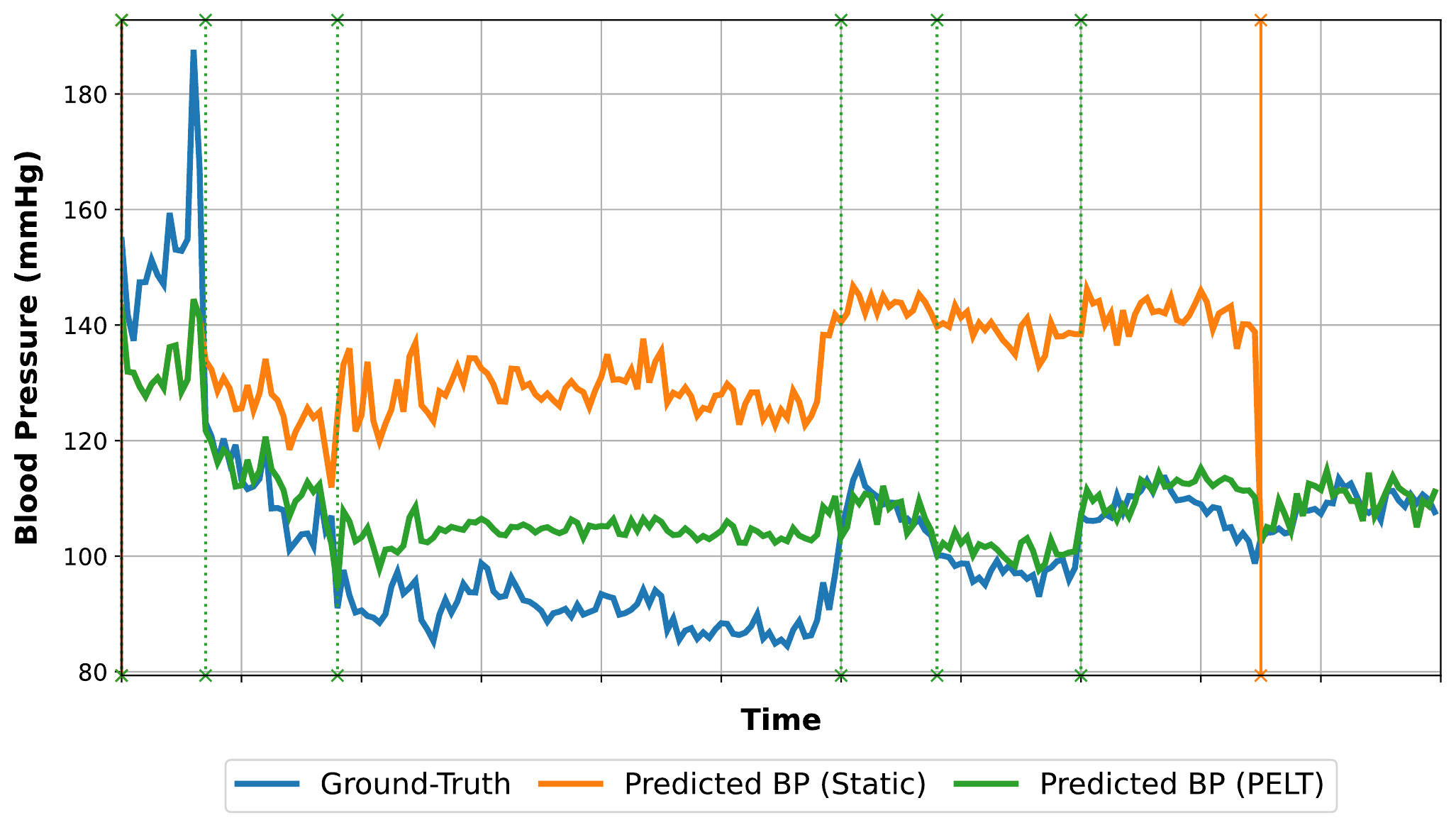}
    \vspace{-0.2in}
    \caption{Qualitative results for calibration-based methods using periodic re-calibration (orange) and targeted re-calibration at BP fluctuation points (green).}
    \label{fig.qual1}
\end{figure*}

\begin{figure*}[t]
    \centering
    \includegraphics[width=1\textwidth]{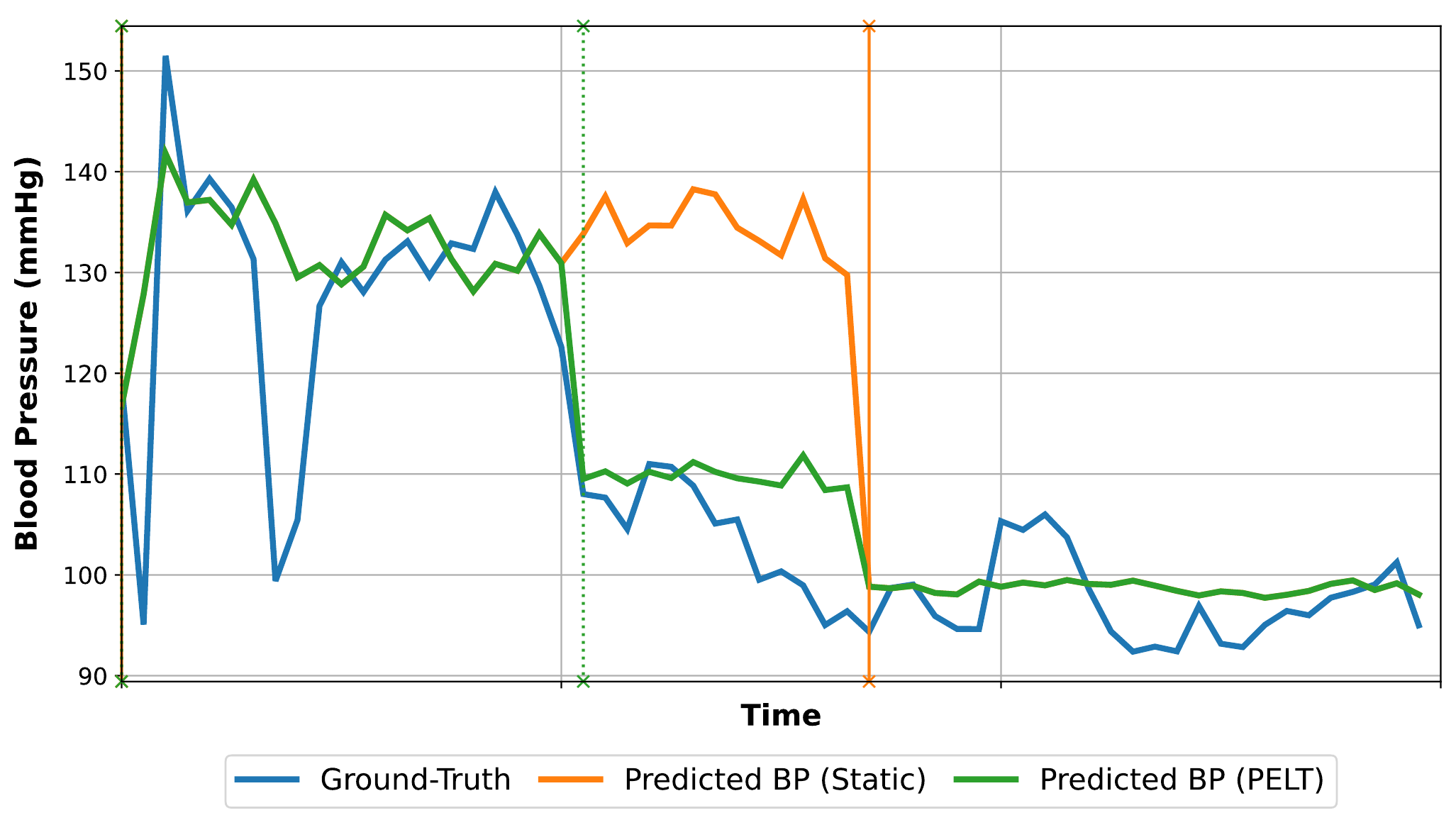}
    \vspace{-0.2in}
    \caption{Qualitative results for calibration-based methods using periodic re-calibration (orange) and targeted re-calibration at BP fluctuation points (green).}
    \label{fig.qual2}
\end{figure*}

\begin{figure*}[t]
    \centering
    \includegraphics[width=1\textwidth]{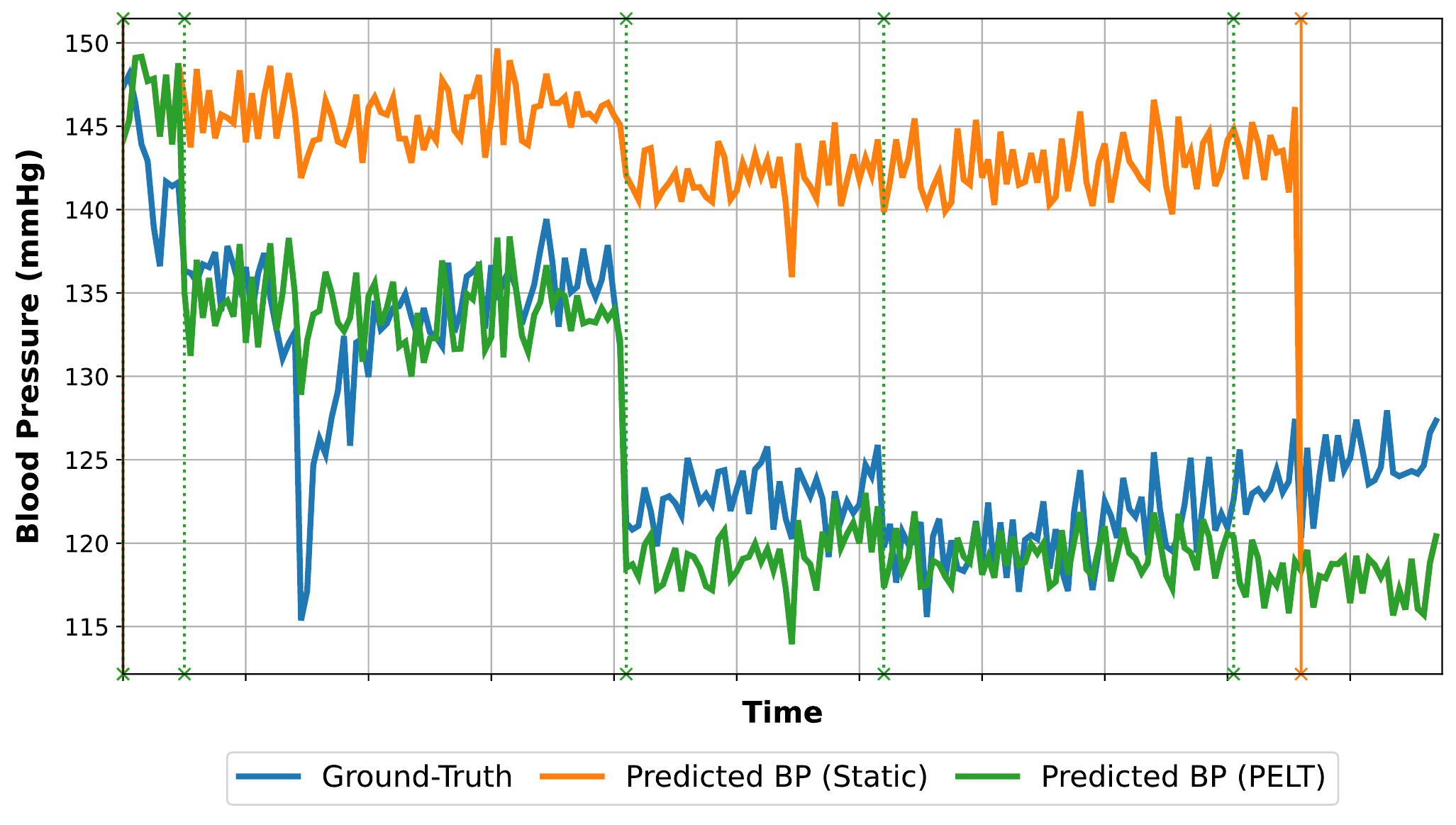}
    \vspace{-0.2in}
    \caption{Qualitative results for calibration-based methods using periodic re-calibration (orange) and targeted re-calibration at BP fluctuation points (green).}
    \label{fig.qual3}
\end{figure*}

Figures~\ref{fig.qual1}--\ref{fig.qual3} show representative examples comparing the periodic calibration (Static) baseline and our PELT-based re-calibration against ground-truth BP. In all cases, the periodic calibration exhibits persistent bias when patient hemodynamics shift over time, failing to adapt to sustained BP changes. The PELT model detects these transitions via change point detection and re-calibrates accordingly, maintaining closer agreement with the ground truth throughout the session.

\section{Sensitivity Analysis of PELT Hyperparameters}

We selected the PELT hyperparameters via greedy search, using the average blood pressure change within detected segments, $\mathrm{Mean}|\Delta_{\mathrm{Seg}}|$, as the selection criterion. This metric favors segmentations that capture meaningful BP fluctuations rather than noise-induced over-segmentation.

The selected default configuration provided the best overall trade-off. The cost model had the largest effect on segmentation behavior: \texttt{l1} and \texttt{l2} produced many more detected points per case but much smaller segment-level BP changes, indicating sensitivity to noise rather than meaningful BP transitions. 
% The penalty parameter also affected segmentation granularity, whereas \texttt{jump} and \texttt{min\_size} showed limited sensitivity in preliminary experiments and are omitted for brevity.

\begin{table}[t]
\centering
\caption{Sensitivity analysis of PELT hyperparameters. The selected default values are marked with an asterisk.}
\label{tab:pelt_sensitivity}
\begin{tabular}{llccc}
\toprule
Parameter & Value & Points/Case & $|\Delta_{\mathrm{Seg}}|$ SBP & $|\Delta_{\mathrm{Seg}}|$ DBP \\
\midrule
Penalty & 1 & 26.75 & 11.69 & 6.16 \\
        & 5$^{*}$ & 9.09 & 13.18 & 6.81 \\
        & 10 & 6.00 & 12.04 & 6.27 \\
\midrule
Model   & \texttt{l1} & 56.72 & 7.52 & 3.94 \\
        & \texttt{l2} & 121.70 & 6.32 & 3.33 \\
        & \texttt{rbf}$^{*}$ & 9.09 & 13.18 & 6.81 \\
        & \texttt{linear} & 50.35 & 6.92 & 3.76 \\
\bottomrule
\end{tabular}
\end{table}

\section{Hyperparameter Sensitivity of the Online CPD Module}
For the online CPD module, hyperparameters were selected based on the intersection-over-union (IoU), which directly reflects the overlap between predicted unstable intervals and the reference intervals. Table~\ref{tab:online_cpd_sensitivity} summarizes the sensitivity analysis for the main hyperparameters. Overall, the results show a clear trade-off between classification accuracy and interval-level localization quality. For $\eta_{\mathrm{base}}$, decreasing the value increased accuracy but reduced IoU, indicating that a more permissive threshold improved point-wise detection at the expense of precise interval matching. For $\lambda$, larger values caused the threshold to tighten too rapidly, which substantially degraded both accuracy and IoU. Based on these results, we selected $\eta_{\mathrm{base}}=0.99$ and $\lambda=0.05$ as the default configuration, as they achieved the best IoU while maintaining stable AUROC and accuracy.

\begin{table}[t]
\centering
\caption{Sensitivity analysis of online CPD hyperparameters. The selected default values are marked with an asterisk.}
\label{tab:online_cpd_sensitivity}
\begin{tabular}{llccc}
\toprule
\textbf{Parameter} & \textbf{Value} & \textbf{AUROC} & \textbf{Accuracy} & \textbf{IoU} \\
\midrule
$\eta_{\mathrm{base}}$ & 0.999   & 0.7271 & 0.6412 & 0.6140 \\
                       & 0.99$^{*}$ & 0.7275 & 0.6457 & 0.6149 \\
                       & 0.9     & 0.7190 & 0.7214 & 0.5656 \\
\midrule
$\lambda$             & 1.0     & 0.7444 & 0.2391 & 0.2793 \\
                      & 0.1     & 0.7401 & 0.5271 & 0.5758 \\
                      & 0.05$^{*}$ & 0.7275 & 0.6457 & 0.6149 \\
\bottomrule
\end{tabular}
\end{table}

\section{Further Evaluations}

\subsection{BP Estimation Performance on Unstable BP Intervals with Targeted Re-Calibration\label{apd:f2}}

Section~\ref{sec6.2} showed that calibration-free models degrade substantially within unstable
BP intervals. Here we examine whether the benefit of targeted re-calibration is retained
under the same conditions. Table~\ref{tab:pelt-unstable} extends Table~\ref{table.three} by reporting
\textbf{PPG2BP-Net} performance separately over the whole recording and over unstable BP
intervals, defined as in Section~\ref{section.3.2}. Every calibration strategy exhibits higher SBP
error within unstable intervals than over the whole recording, confirming that these
segments remain the most difficult regions to track. The relative ordering of the
strategies is nevertheless preserved: periodic calibration alone yields 10.93 mmHg SBP
MAE, the \textit{$\Delta$BP} and \textit{Random} strategies reduce the error to 9.46 and 9.10 mmHg,
respectively, and \textit{PELT} achieves 6.73 mmHg.
Notably, \textit{PELT} also shows the smallest gap between whole-interval and unstable-interval
performance (0.89 mmHg, compared with 1.41--1.66 mmHg for the remaining strategies),
indicating that aligning re-calibration timing with distributional shifts is most
effective precisely where estimation is hardest. For DBP, the differences across
strategies are smaller, and the \textit{$\Delta$BP} criterion attains the lowest error within
unstable intervals (3.88 mmHg) despite requiring nearly twice as many calibration points
per case; this is consistent with our definition of unstable intervals, which is driven
by systolic and mean arterial pressure levels and therefore couples less directly to
diastolic dynamics. Overall, these results indicate that the advantage of targeted
re-calibration reported in Section~\ref{sec6.2} is not an artifact of averaging over stable
periods, but is retained—and for SBP amplified—under clinically critical hemodynamic
conditions.

\begin{table}[t]
\centering
  \caption{Impact of targeted re-calibration using ground-truth (offline) blood pressure change points, evaluated with \textbf{PPG2BP-Net}. Mean absolute error in mmHg with 95\% confidence intervals are reported over the whole recording (\textit{Whole}) and over unstable BP intervals defined in Section~\ref{section.3.2}. Points / Case denotes the average number of re-calibration points per case (mean ± std).
  The plus sign (``$+$'') keeps the 120-minute schedule as the base and adds the change points of strategies such as \textit{$\Delta$BP}, \textit{Random}, and \textit{PELT}. \textit{Random} uses the per-case point count of PELT with uniformly sampled locations.}
  \resizebox{\columnwidth}{!}{%
  \begin{tabular}{lccccc}
  \toprule
  \multirow{2}{*}{\textbf{Calibration Points}} &
  \multicolumn{2}{c}{\textbf{Whole}} &
  \multicolumn{2}{c}{\textbf{Unstable BP Intervals}} &
  \multirow{2}{*}{\textbf{Points / Case}} \\
  \cmidrule(lr){2-3} \cmidrule(lr){4-5}
   &
  \textbf{SBP  ↓} & \textbf{DBP  ↓} &
  \textbf{SBP  ↓} & \textbf{DBP  ↓} & 
   \\
  \midrule
  Every 120 minutes          & 9.33 ± 0.05 & 5.21 ± 0.03 & 10.93 ± 0.17& 5.78 ± 0.09
& 3.30 ± 1.52 \\ \midrule
  \hphantom{/}+ \textit{$\Delta$BP}       & 8.05 ± 0.05          & 4.53 ± 0.03         & 9.46 ± 0.15& \textbf{3.88 ± 0.08}& 17.36 ± 22.57 \\
  \hphantom{/}+ \textit{Random}      & 7.44 ± 0.04          & 4.01 ± 0.03          & 9.10 ± 0.15& 4.63 ± 0.09
& 9.91 ± 4.02 \\
  \hphantom{/}+ \textit{PELT}       & \textbf{5.84 ± 0.04} & \textbf{3.85 ± 0.02} & \textbf{6.73 ± 0.13}& 4.11 ± 0.07& 9.82 ± 4.02 \\
  \bottomrule
  \end{tabular}}
\label{tab:pelt-unstable}
\end{table}

\subsection{Classification Baseline for CPD\label{apd:f1}}

As no prior work formulates online BP change point detection from PPG signals alone,
a directly comparable baseline is not available for the results reported in Table~\ref{table.four}.
To place these numbers in context, we additionally train a naive binary classifier that
predicts, for each incoming 10-second PPG segment, whether the segment corresponds to a
BP change point, supervised by the same PELT-derived change point labels. The classifier
follows the encoder configuration and training protocol described in
Appendix~\ref{apd:third}, and replaces the supervised contrastive objective in
Section~\ref{section.4.2.1} and the sequential clustering procedure in Section~\ref{section.4.2.2} with a binary
cross-entropy loss. The predicted change points are then used to trigger targeted
re-calibration with \textbf{PPG2BP-Net} under the same protocol as Table~\ref{table.five}. Table~\ref{tab:cpd_w_cls} reports the
comparison for two backbones, Inception1D and PaPaGei-S. For both backbones, the
classification baseline remains close to chance level in AUROC (0.5332 and 0.5321),
whereas the proposed method achieves 0.7275 and 0.7223, with the same ordering
observed in IoU. The detection gap is likewise reflected in the downstream task: SBP MAE
decreases from 8.31 to 7.27 mmHg with Inception1D and from 7.79 to 7.30 mmHg with
PaPaGei-S, while DBP MAE decreases from 4.56 and 4.13 mmHg to 3.87 mmHg in both cases.
Detection quality thus tracks the gain obtained from targeted re-calibration, and the
consistency of this pattern across a from-scratch and a pretrained backbone indicates
that the improvement originates from the detection formulation rather than from the
choice of encoder.

\begin{table}[t]
\centering
\small
\caption{Comparison with a naive binary classification baseline for online change point
detection. Detection performance and the corresponding downstream BP estimation error
using \textbf{PPG2BP-Net} are reported.}
\label{tab:cpd_w_cls}
\begin{tabular}{llcccc}
\toprule
\multirow{2}{*}{\textbf{CPD Backbone}}& \multirow{2}{*}{\textbf{CPD Method}}& \multicolumn{2}{c}{\textbf{CPD}} & \multicolumn{2}{c}{\textbf{BP Estimation}}\\
\cmidrule(lr){3-4} \cmidrule(lr){5-6}
& & \textbf{AUROC  ↑}& \textbf{IoU  ↑}& \textbf{SBP MAE  ↓} &\textbf{DBP MAE  ↓}\\
\midrule
\multirow{2}{*}{Inception1D}& Classification& 0.5332& 0.3919& 8.31 ± 0.04&4.56 ± 0.03\\
                       & Ours& 0.7275& 0.6149& 7.27 ± 0.04&3.87 ± 0.02\\
\midrule
\multirow{2}{*}{PaPaGei-S}& Classification& 0.5321& 0.4552& 7.79 ± 0.04&4.13 ± 0.02\\
                      & Ours& 0.7223& 0.6062& 7.30 ± 0.04&3.87 ± 0.02\\
\bottomrule
\end{tabular}
\label{tab:11}
\end{table}

\subsection{Online CPD-Triggered Re-Calibration on Unstable BP Intervals}
\label{app:online-unstable}

Table~\ref{tab:onlinecpd-unstable} extends Table~\ref{table.five} by reporting \textbf{PPG2BP-Net} performance
separately over the whole recording and over the unstable BP intervals defined in
Section~\ref{section.3.2}, and additionally includes the classification baseline described in
Appendix~\ref{apd:f1} as a trigger. The ordering observed over the whole
recording is preserved within unstable intervals: periodic calibration alone yields
10.93 mmHg SBP MAE, the \textit{Classification} and \textit{Random} triggers reduce the error to 8.73 and
8.06 mmHg, and \textit{Online CPD} achieves the lowest error at 7.76 mmHg, with the same ordering
in DBP. Two comparisons isolate the source of this gain. First, \textit{Online CPD} and \textit{Random}
issue an almost identical number of re-calibration points per case (8.77 for both), so
the remaining difference is attributable to when re-calibration is applied rather than
how often. Second, the \textit{Classification} trigger performs worse than both despite issuing
substantially more points (12.31 per case), indicating that inaccurately localized change
points do not compensate for their higher frequency. All triggered strategies also
exhibit a smaller degradation from whole-interval to unstable-interval performance
(0.42--0.49 mmHg SBP) than periodic calibration alone (1.60 mmHg), confirming that the
benefit of targeted re-calibration is retained under clinically critical hemodynamic
conditions when change points are detected online.

\begin{table}[t]
\centering
  \caption{Performance of targeted re-calibration using the proposed online change point detection. Mean absolute error in mmHg with 95\% confidence intervals are reported. Points / Case denotes the average number of re-calibration points per case (mean ± std).
  The plus sign (``$+$'') keeps the 120-minute schedule as the base and adds the change points of strategies such as \textit{Random}, \textit{Classification}, and \textit{Online CPD}. \textit{Random} uses the per-case point count of \textit{Online CPD} with uniformly sampled locations. Both \textit{Classification} and \textit{Online CPD} use Inception1D as the CPD backbone.
  }
  \resizebox{\columnwidth}{!}{%
  \begin{tabular}{lccccc}
  \toprule
  \multirow{2}{*}{\textbf{Calibration Points}} &
  \multicolumn{2}{c}{\textbf{Whole}} &
  \multicolumn{2}{c}{\textbf{Unstable BP Intervals}} &
  \multirow{2}{*}{\textbf{Points / Case}} \\
  \cmidrule(lr){2-3} \cmidrule(lr){4-5}
   &
  \textbf{SBP  ↓} & \textbf{DBP  ↓} &
  \textbf{SBP  ↓} & \textbf{DBP  ↓} & 
   \\
  \midrule
  Every 120 minutes          & 9.33 ± 0.05 & 5.21 ± 0.03 & 10.93 ± 0.17& 5.78 ± 0.09
& 3.30 ± 1.52 \\
  \midrule
  \hphantom{/}+ \textit{Classification} & 8.31 ± 0.04& 4.56 ± 0.03& 8.73 ± 0.09& 4.72 ± 0.05
& 12.31 ± 22.59\\
  \hphantom{/}+ \textit{Random}         & 7.60 ± 0.02& 4.07 ± 0.01& 8.06 ± 0.09& 4.33 ± 0.05
& 8.77 ± 3.34
\\
  \hphantom{/}+ \textit{Online CPD}     & \textbf{7.27 ± 0.02}& \textbf{3.87 ± 0.01}& \textbf{7.76 ± 0.08}& \textbf{4.15 ± 0.05}& 8.77 ± 3.33\\
  \bottomrule
  \end{tabular}}
\label{tab:onlinecpd-unstable}
\end{table}

\subsection{Computational Cost}
Table~\ref{tab:latency} reports the computational cost of the online CPD module. At
inference, the pipeline performs a single encoder forward pass per incoming 10-second
PPG segment, followed by cosine similarity comparisons against the stored centroids,
which require $O(K|\mathcal{C}|d)$ operations, where $d$ is the feature dimension, $K$
the buffer size, and $|\mathcal{C}|$ the number of centroids. Across all backbones, the
forward pass takes 9.2--31.0 ms while the sequential clustering step remains below
0.05 ms. Since a prediction is required only once every 10 seconds, the end-to-end
latency leaves a large margin for real-time operation, and the overall cost is dominated
by the encoder, with the proposed detection procedure adding negligible overhead on top
of the backbone already used for BP estimation.

\begin{table}[t]
\centering
\caption{Analysis on computation cost and latency of online change point detection}
    \begin{tabular}{lcccc}
    \toprule
    \textbf{CPD Backbone}& \textbf{Params (M)}& \textbf{FLOPs (M)}& \textbf{Forward (ms)}& \textbf{Cluster (ms)}\\
    \midrule
    Inception1D & 0.5
& 591.8
& 21.9& 0.03
\\
    PaPaGei-S & 5.8
& 76.6
& 9.2& 0.03
\\
    Pulse-PPG & 28.5
& 1140.5
& 31& 0.03
\\
    AnyPPG & 5.9& 245.5& 26.2& 0.04\\
    \bottomrule
    \end{tabular}
\label{tab:latency}
\end{table}

\section{Clinical Interpretation of Detected Change Points}
\label{app:clinical-interpretation}

\paragraph{What transitions PELT captures.} To characterize the physiological events
underlying the detected change points, we cross-reference the unstable BP intervals
defined in Section~\ref{section.3.2} with vasopressor administration records available in VitalDB.
As shown in Table~\ref{tab:vasopressor}, the large majority of unstable segments
(6,094 segments, 85.4\%) occur while a vasopressor is being administered, compared with
1,038 segments (14.6\%) in its absence. This indicates that the transitions identified by
PELT are not arbitrary statistical fluctuations but concentrate around pharmacologically
driven hemodynamic shifts, which are precisely the events during which a stale
calibration reference is most likely to mislead the estimation model. This also supports
the use of change point-triggered re-calibration as a clinically meaningful, rather than
purely data-driven, criterion.

\paragraph{Clinical consequences in unstable intervals.} The practical relevance of the
residual error in these intervals follows from where the intervals are located. Under
our definition, unstable segments lie near the SBP $>$ 180 mmHg and MAP $<$ 65 mmHg
boundaries, and the SBP error observed within them remains approximately 11--14 mmHg for
calibration-based models (Table~\ref{table.three}). An error of this magnitude at these boundaries may
delay the recognition of a hypertensive emergency, for which timely treatment is
recommended in the acute care setting~\citep{bress2024management}, or the initiation of vasopressor
and fluid therapy in the management of sepsis and septic shock~\citep{evans2021surviving}. This
motivates reporting estimation accuracy within fluctuation intervals separately, rather
than relying on whole-recording averages that mask errors in exactly these regions.

\begin{table}[t]
\centering
\caption{Distribution of unstable BP segments in VitalDB with respect to vasopressor
administration.}
    \begin{tabular}{lc}
    \toprule
    \textbf{Context} & \textbf{Unstable Segments (\%)} \\
    \midrule
    Vasopressor    & 6,094 (85.4\%) \\
    No vasopressor & 1,038 (14.6\%) \\
    \bottomrule
    \end{tabular}
\label{tab:vasopressor}
\end{table}

\end{document}